**NetSciEd: Network Science and Education for the Interconnected World**

Hiroki Sayama[1*,2,3], Catherine Cramer[4], Lori Sheetz[5], and Stephen Uzzo[4]

[1] Center for Collective Dynamics of Complex Systems / Department of Systems Science and Industrial Engineering, Binghamton University, State University of New York, P.O. Box 6000, Binghamton, NY 13902-6000 (*corresponding author: sayama@binghamton.edu)

[2] Center for Complex Network Research, Northeastern University, 177 Huntington Ave., 11th floor, Boston, MA 02115

[3] School of Commerce, Waseda University, 1-104 Totsukamachi, Shinjuku-ku, Tokyo 169-8050, Japan

[4] New York Hall of Science, 47-01 111th St., Corona, NY, 11368

[5] Center for Leadership and Diversity in S.T.E.M., United States Military Academy, 601 Cullum Rd., Thayer Hall Room 240, West Point, NY 10996

## Abstract

This short article presents a summary of the NetSciEd (Network Science and Education) initiative that aims to address the need for curricula, resources, accessible materials, and tools for introducing K-12 students and the general public to the concept of *networks*, a crucial framework in understanding complexity. NetSciEd activities include (1) the NetSci High educational outreach program (since 2010), which connects high school students and their teachers with regional university research labs and provides them with the opportunity to work on network science research projects; (2) the NetSciEd symposium series (since 2012), which brings network science researchers and educators together to discuss how network science can help and be integrated into formal and informal education; and (3) the Network Literacy: Essential Concepts and Core Ideas booklet (since 2014), which was created collaboratively and subsequently translated into 18 languages by an extensive group of network science researchers and educators worldwide.

## Introduction

Since its boom in the late 20th century (Watts & Strogatz 1998; Barabási & Albert 1999), the concept of *networks*, a crucial framework in understanding complexity, has become evermore relevant to people's everyday life (Buchanan 2003; Newman 2010; Caldarelli & Catanzaro 2012; Barabási 2014). Nearly every system that surrounds us, or exists among or inside of us, can be understood as a network, i.e., a discrete structure that consists of nodes (vertices, entities, actors, items) and the edges (links, relationships, ties, connections) that connect the nodes. Examples of networks are ubiquitous, including the Internet, social media, financial systems, transportation networks, ecosystems, organizations, friendships, schools, classrooms, learning materials, brains, immune systems, and even genes/proteins within a single cell. Knowledge about networks can help us to make sense of those systems, making it a useful literacy for people to be effective and successful in this increasingly complex, interconnected world of the 21st century (Sheetz, Dunham, & Cooper 2015; Sayama et al. 2016). Network science (Barabási 2013), an interdisciplinary field of scientific research on networks, offers a powerful approach for conceptualizing, developing, and understanding solutions to complex social, health, technological, and environmental problems.

While network science can provide opportunities to develop many of the skills, habits of mind, and core ideas that are highly relevant to today's interconnected world (Harrington et al. 2013; Sánchez & Brändle 2014), they are neither addressed nor utilized in extant elementary/secondary education curricula and teaching practice. Current education systems are still based predominantly on reductionistic mindsets, in which teaching is conducted on a subject-by-subject and module-by-module basis and improvement is planned and implemented using a linear, causal, independent problem-to-solution approach, with very little consideration given to the nontrivial interconnectedness among various factors and components involved in these complex systems. We believe that this situation presents an opportunity and a need for a multifaceted intervention at community levels, which should be by itself a coordinated network of collaborative efforts, including the development of curricular materials and resources; promoting network thinking among K-12 students, teachers, and administrators; educating educators and the general public about networks; and increasing the awareness of the demand for network science education among researchers, as well as many other possible activities. Theories of complex systems inform us of the possibility of a "phase transition" induced by such collective actions taking place on a dynamical network of interconnected components. Such a scenario of cascading, global change is what is envisioned in all the activities described below.

In this short article, we present a summary of the NetSciEd (Network Science and Education; http://tinyurl.com/netscied) initiative that we have been running over the last several years to address the educational need described above. NetSciEd activities include (1) the NetSci High educational outreach program (Cramer et al. 2015b; http://bit.ly/2qwS3oh), which connects high school students and their teachers with regional university research labs and provides them with the opportunity to work on network science research projects, (2) the NetSciEd symposium series (http://bit.ly/2cNd7RV), which brings network science researchers and educators together to discuss how network science can help and be integrated into school education, and (3) the Network Literacy: Essential Concepts and Core Ideas booklet (Cramer et al. 2015a; http://tinyurl.com/networkliteracy), which was created collaboratively and subsequently translated into 18 languages by an extensive group of network science researchers and educators worldwide.

**NetSci High**

NetSci High (Cramer et al. 2015b; http://bit.ly/2qwS3oh) is a high school educational outreach program that brings network science to high schools through workshops and yearlong, open-ended research projects supervised by graduate students, postdocs, and faculty at regional university research labs. One of the objectives of NetSci High is to provide high school students and teachers an interdisciplinary learning experience that transcends the boundaries among traditional academic subjects (e.g., languages, mathematics, science, social studies, art, music, etc.). The concept of networks can be easily understood and applied to various subjects, which is ideal for achieving such an interdisciplinary educational objective.

Since its inception as a small pilot project created by the New York Hall of Science and Binghamton University in 2010, NetSci High has taken several different formats with various funding sources (see Table 1 for an overview). Its first year (2010-2011) was run as a poster competition among seven high school teams in the New York City, Boston, and Binghamton areas. In the second year (2011-2012), it took the form of travel scholarships for two high school teams from the Binghamton area. Supplemental funding from the U.S. National Science Foundation (NSF) and a corporate donation from BAE Systems supported these first two years of activities.

In the subsequent three years (2012-2015), NetSci High was further developed as an NSF ITEST (Innovative Technology Experiences for Students and Teachers) project "Network Science for the Next Generation," with Boston University and the United States Military Academy at West Point as official partner institutions, and Newburgh, NY, was added as an additional target geographical region. Two-week summer workshops were developed as an initial kick-off event for the students' yearlong research experience (Figure 1). Several educational modules and activities were developed for these summer workshops (for teachers as well as for the students) (http://bit.ly/2rfGWB3), which were highly effective in motivating and fully preparing students and teachers for network science research projects.

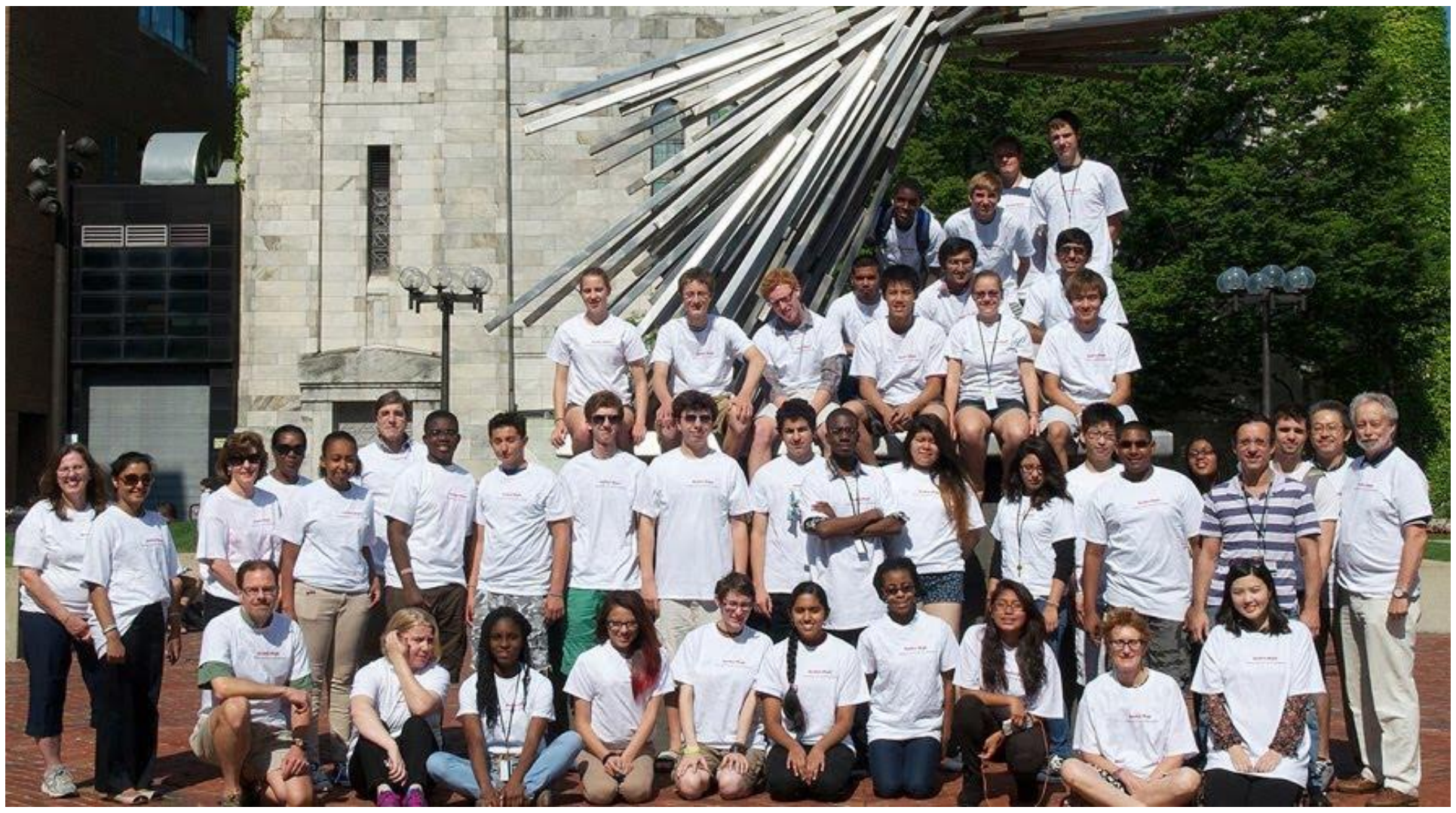

Figure 1: Participants at the NetSci High summer workshop at Boston University in July 2013.

At the end of each academic year, NetSci High student teams had an opportunity to attend academic conferences and present their work themselves as the final output of their research, depending on the availability of appropriate conferences and travel funding. Several student teams also published their work on their own outside the NetSci High program (Blansky et al. 2013; Onwuka, Pineda, & Stepakoff 2015; Shah et al. 2015; Asante, McMillan, & Peticco 2016; Chen & Mendoza 2016; Mariano 2016).

Table 1: Overview of NetSci High educational outreach program.
[*] NetSci: International School and Conference on Network Science; [**] ICCS: International Conference on Complex Systems;
[***] CompleNet: International Conference on Complex Networks

| Year | Format | Participating schools and # of students participated | Participating universities | Final conference(s) |
|---|---|---|---|---|
| **2010-2011** | Poster competition | Flushing International High School (Flushing, NY) – 2<br>Elmont Memorial High School (Elmont, NY) – 1<br>Thomas A. Edison Career and Technical Education High School (Jamaica, NY) – 1<br>Jericho High School (Jericho, NY) – 1<br>Maine-Endwell High School (Endwell, NY) – 7<br>Windsor School (Boston, MA) – 2 | City College of New York<br>Columbia University<br>St. John's University<br>Binghamton University<br>Harvard Medical School | NetSci[*] 2011, ICCS[**] 2011 |
| **2011-2012** | Travel scholarship | Maine-Endwell High School (Endwell, NY) – 4<br>Vestal High School (Vestal, NY) – 2 | Binghamton University | NetSci[*] 2012 |
| **2012-2013** | Workshop + supervised research | Boston University Academy (Boston, MA) – 7<br>Newburgh Free Academy (Newburgh, NY) – 7<br>Vestal High School (Vestal, NY) – 2 | Boston University<br>Harvard Medical School<br>US Military Academy<br>Binghamton University | NetSci High Conference at Boston University |
| **2013-2014** | Workshop + supervised research | Boston University Academy (Boston, MA) – 4<br>Chelsea High School (Chelsea, NY) – 4<br>Elmont Memorial High School (Elmont, NY) – 6<br>Newburgh Free Academy (Newburgh, NY) – 7<br>Vestal High School (Vestal, NY) – 8 | Boston University<br>Harvard Medical School<br>Stevens Institute of Technology<br>Columbia University<br>US Military Academy<br>Binghamton University | NetSci[*] 2014, NetSci High Conference at Boston University |
| **2014-2015** | Workshop + supervised research | CATS Academy (Braintree, MA) – 3<br>Chelsea High School (Chelsea, NY) – 4<br>Newburgh Free Academy (Newburgh, NY) – 12<br>Vestal High School (Vestal, NY) – 8 | Harvard Medical School<br>Stevens Institute of Technology<br>US Military Academy<br>Binghamton University | CompleNet[***] 2015 |

The summative assessments of NetSci High (Faux 2015) have shown that the program had a significant impact on students' understanding of networks and their importance, and on the development of students' basic literacy of data driven science (e.g., data collection, manipulation, analysis; mathematical/computational skills; visualization), academic and scientific skills (e.g., creative/critical thinking; hypothesis forming; technical writing and presentation; interdisciplinary interest), and professional skills (e.g., time management; collaborative skills; self-confidence; leadership). These constitute an excellent preparation for high school students for higher education and Science/Technology/Engineering/Math (STEM)-related career paths.

Another important accomplishment of NetSci High is that it educated not just high school students and teachers but also the scientific community in return. Continuous participation of high school student teams at network science conferences has increased the awareness of educational values of network science, as well as the academic potential of students in secondary education, among the researchers in the network science community. As a result, education is now being recognized as one of the key themes in network science and related complexity fields. While federal financial support officially ended in 2015, the NetSci High model of educational outreach has made a lasting impact and has been adopted in several other outreach programs (e.g., Cramer 2015; Trunfio 2016; Mangal et al. 2016; United States Military Academy Center for Leadership and Diversity in S.T.E.M. 2017). The resources of the summer workshops are made publicly available for others to replicate the program (http://bit.ly/2rfGWB3).

**NetSciEd Symposium Series**

Concurrently with the NetSci High program described above, we have been organizing the NetSciEd symposium series at NetSci conferences every year since 2012 (http://bit.ly/2cNd7RV; Figure 2). These symposia aim to bring network science researchers and educators together to discuss any issues that arise at the intersection between network science and education. Symposium programs typically include oral presentations, posters, and roundtable discussions. NetSci High student teams' posters have also been displayed at the symposium venues.

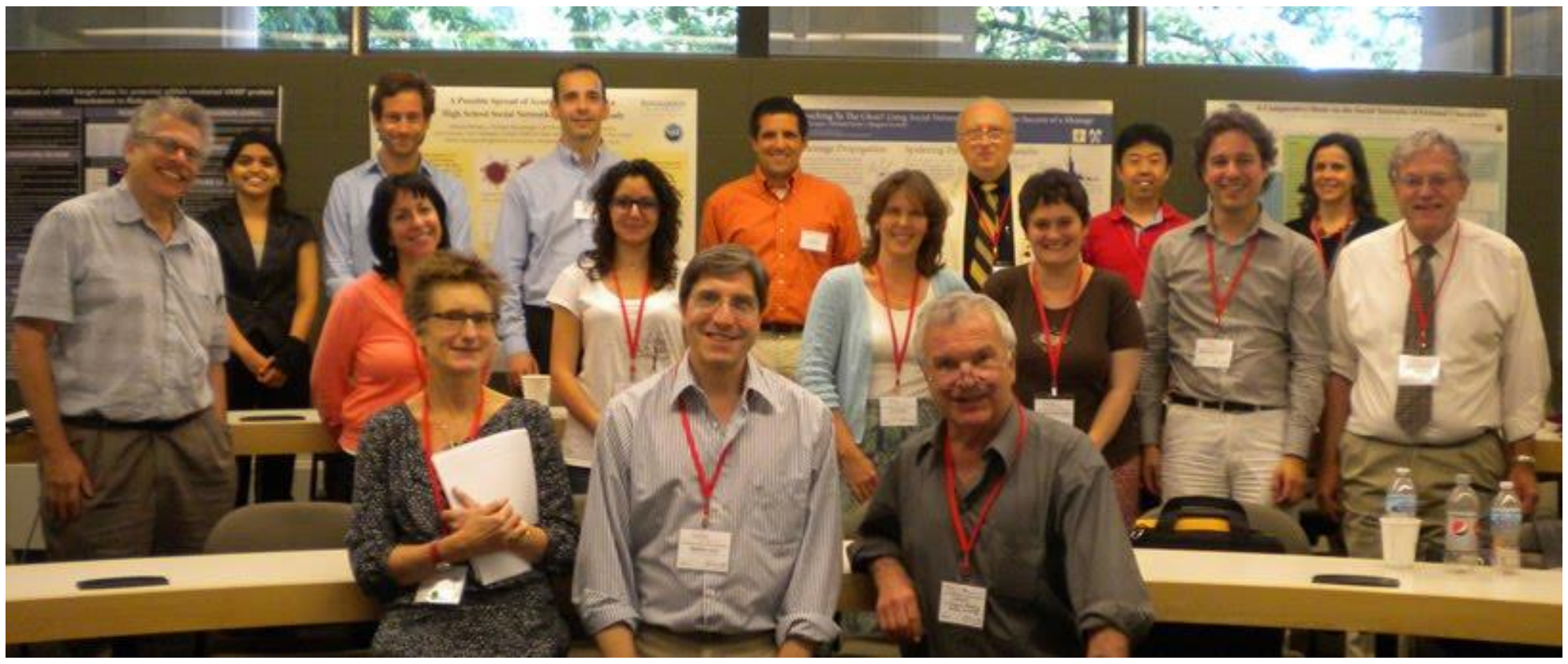

Figure 2: Participants at the First NetSciEd symposium in Evanston, IL, in June 2012.

Table 2 shows basic statistics of the NetSciEd symposium series. While the symposium programs were typically made of solicited talks in the first few years, the number of unsolicited abstract submissions has been rapidly increasing over the last few years, indicating the growing interest in education as a research topic among network science researchers.

Table 2: Basic statistics of NetSciEd symposia.

| Name | Year | Location | Total # of presentations | # of unsolicited submissions | Format |
|---|---|---|---|---|---|
| **NetSciEd** | 2012 | Evanston, IL | 9 | 0 | Half-day |
| **NetSciEd2** | 2013 | Copenhagen, Denmark | 11 | 0 | Half-day |
| **NetSciEd3** | 2014 | Berkeley, CA | 10 | 0 | Half-day |
| **NetSciEd4** | 2015 | Zaragoza, Spain | 12 | 1 | Full-day |
| **NetSciEd5** | 2016 | Seoul, Korea | 8 | 2 | Half-day |
| **NetSciEd6** | 2017 | Indianapolis, IN | 15 | 6 | Half-day |

Topics discussed at NetSciEd symposia include: (1) educational outreach activities, tools, and materials about network science; (2) curricular development and practices for teaching network science at schools/colleges; (3) teacher education and informal education about networks; (4) network modeling and analysis of educational systems, curricular materials, or classroom/school dynamics; (5) network modeling and analysis of learning processes; and (6) applications of network science for the improvement of education. Table 3 shows an informal breakdown of topics presented at the NetSciEd symposium each year. While the first two topics consistently attract the largest attention of the presenters, there has been a considerable diversity of topics being discussed.

Table 3: Informal breakdown of topics presented at NetSciEd symposia.

| Topic | 2012 | 2013 | 2014 | 2015 | 2016 | 2017 |
|---|---|---|---|---|---|---|
| **Outreach, tools, materials** | 4 | 4 | 3 | 8 | 2 | 4 |
| **Curricular development and practice** | 1 | 2 | 2 | 3 | 5 | 4 |
| **Teacher education/informal education** | | 1 | 1 | | | 1 |
| **Modeling/analysis of educational systems** | 2 | 3 | 2 | | 1 | 3 |
| **Modeling/analysis of learning processes** | | 1 | 1 | | | 2 |
| **Application for improving education** | 1 | | 1 | 1 | | 1 |
| **Other** | 1 | | | | | |

Given these statistical trends, it is anticipated that network science and education will continue to be a steadily growing area of research with diversifying topics. One tangible outcome of the NetSciEd symposium series is the forthcoming Springer book "Network Science in Education – Tools and Techniques for Transforming Teaching and Learning" (Cramer et al. 2017). This volume will be a collection of selected contributions presented at the past NetSciEd symposia, summarizing the current state of research and practice in this area and illustrating its potential future directions.

## Network Literacy: Essential Concepts and Core Ideas

One critical missing piece that was repeatedly recognized throughout the NetSci High program and the NetSciEd symposia was the lack of systematic, accessible, easy-to-understand educational materials about networks and their implications for our everyday life, similar to grade-school-level textbooks and workbooks. Some academic communities have successfully produced such educational materials, such as textbooks produced by the System Dynamics community (Quaden, Ticotsky, & Lyneis 2005; Potash & Heinbokel 2011; Anderson & LaVigne 2014) and the "Ocean Literacy" and other science literacy booklets produced by several geoscience communities (http://bit.ly/2qvsw2B). Following the "Ocean Literacy" case (Cava et al. 2005; Schoedinger, Cava, & Jewell 2006; Strang, DeCharon, & Schoedinger 2007) as an example, a yearlong collaborative process was launched in June 2014 to produce "Network Literacy: Essential Concepts and Core Ideas," (Cramer et al. 2015a; http://tinyurl.com/networkliteracy) a free online booklet that summarizes in layman's terms what every person who lives in the 21st century should know about networks by the time he/she graduates from high school.

The production process of the Network Literacy booklet started with a one-day preconference event held before the NetSci 2014 conference in Berkeley, CA in June 2014, at which several network science researchers and educators held a brainstorming and discussion session to create a list of essential concepts and their clusters (Sayama et al. 2016). Similar brainstorming sessions were also held with NetSci High students and teachers to receive their inputs, which later turned out to be extremely helpful and important. Network science was actively utilized in this process, not only as the subject of the booklet, but also as the technical toolbox for analyzing and organizing the generated concepts. Specifically, the generated concepts were connected to each other by the NetSci High students and teachers according to their perceived relatedness, forming a "multigraph" of the concepts (i.e., a network graph in which a pair of concepts can be connected by more than one edge). Then, a modularity-maximizing community detection algorithm that is frequently used in network science (Blondel et al. 2008) was applied hierarchically to this multigraph of concepts to detect important concept clusters. Those clusters were further reviewed and refined by network science researchers. This eventually resulted in the following seven essential concepts:

1. ***Networks are everywhere.***
2. ***Networks describe how things connect and interact.***
3. ***Networks can help reveal patterns.***
4. ***Visualizations can help provide an understanding of networks.***

5. ***Today's computer technology allows you to study real-world networks.***
6. ***Networks help you to compare a wide variety of systems.***
7. ***The structure of a network can influence its state and vice versa.***

For each of these seven essential concepts, a more detailed description with several related core ideas were collaboratively written, and were then revised again collaboratively by the community of network science researchers worldwide through several iterations. After being illustrated with professionally designed visuals and layout (Figure 3), the booklet was published online in March 2015 (http://tinyurl.com/networkliteracy), and hard copies were disseminated at the CompleNet 2015 conference, at which the final cohort of NetSci High students were also presenting their research work. Sayama et al. (2016) provide more details of the complete production process of the Network Literacy booklet.

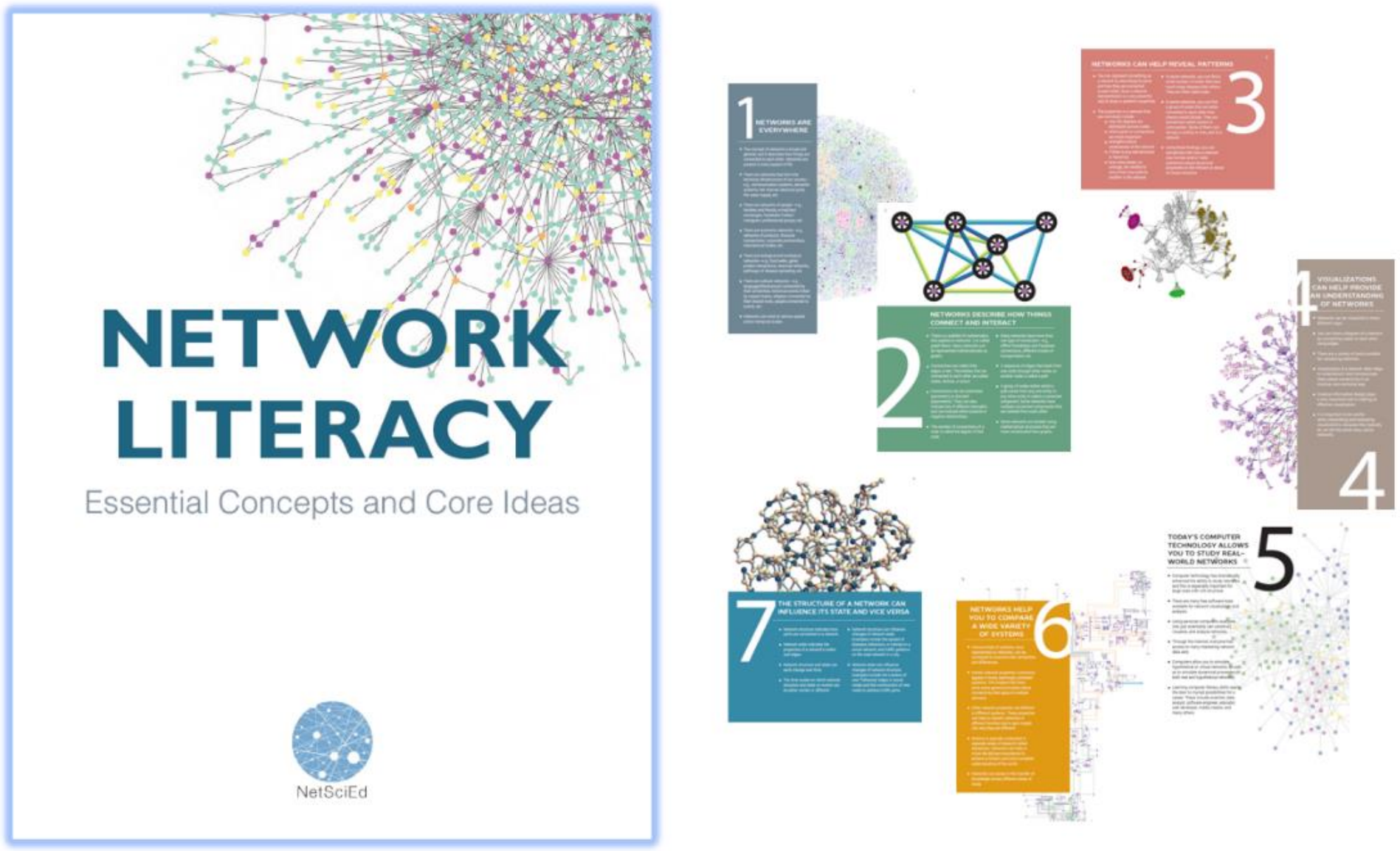

Figure 3: Network Literacy: Essential Concepts and Core Ideas booklet (original English version). Left: cover, right: content page design (design by Eri Yamamoto).

The online release of the Network Literacy booklet was received with overwhelmingly positive responses from researchers and educators worldwide. One notable reaction to this booklet was the coordinated effort to translate its contents into various non-English languages. The translations have been done entirely on a volunteer basis by network science researchers, students, and practitioners who appreciate the importance of learning and understanding networks for the general public at global scales. As of May 2017, the Network Literacy booklet has been translated into the following 18 languages: Arabic, Brazilian Portuguese, Catalan, Chinese (Mandarin), Chinese (traditional Mandarin), French, Dutch, German, Hebrew, Hungarian, Italian, Japanese, Korean, Persian, Polish, Russian, Spanish, and Ukrainian (see Figure 4 for geographical coverage).

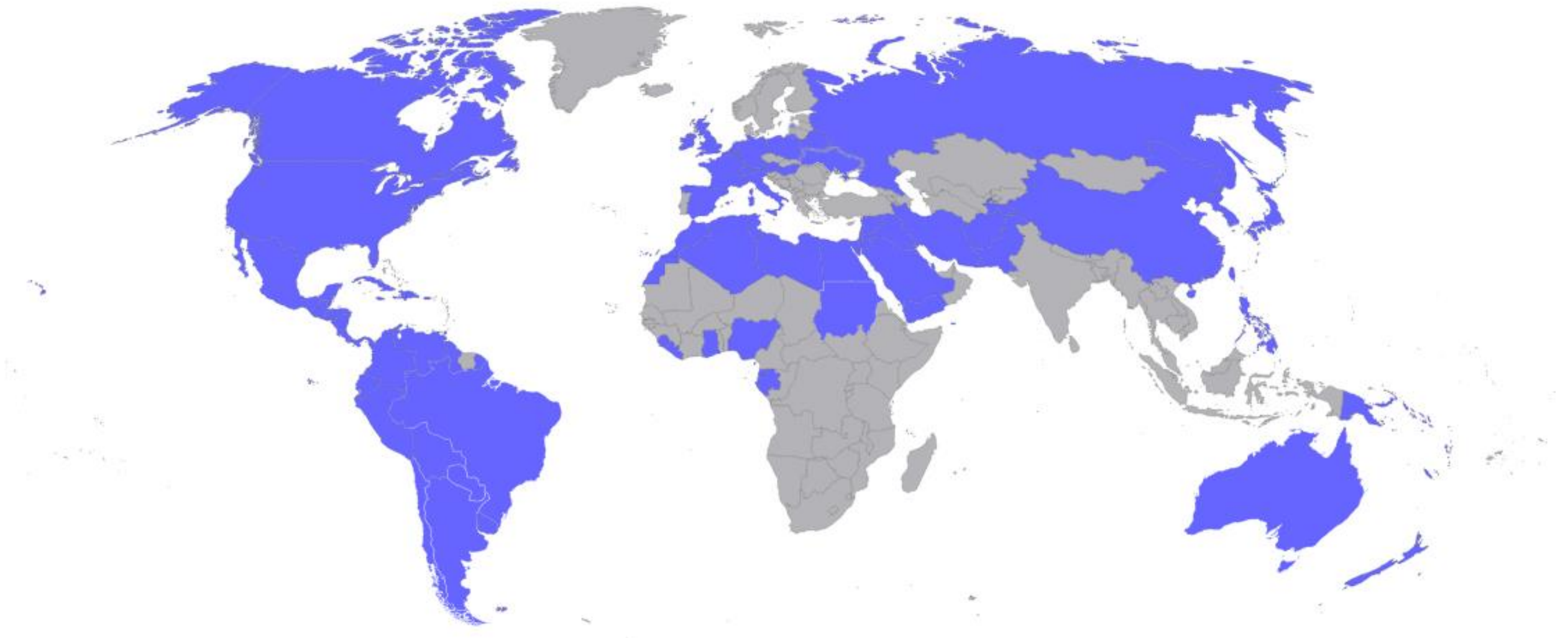

Figure 4: World map of language coverage of the Network Literacy booklet. Network Literacy has been translated into the majority languages of the countries/regions in blue. Chart created using http://www.amcharts.com/visited_countries/.

**Conclusions**

In this article, we presented brief summaries of three different, yet interrelated activities that promote fruitful integration of network science and education. Conducted under the moniker of NetSciEd, these activities are based on our strong belief that learning about and understanding networks has a direct benefit as a basic literacy for everyone who lives in today's highly interconnected world. In the meantime, network science can also offer more immediate, practical benefits, particularly in educational settings. First, network science can offer an accessible conceptual framework that naturally crosses academic subjects and thus helps students develop interdisciplinary academic/scientific/professional skills early on. Second, network science can also offer theoretical/technical tools to model, analyze, and understand the structure and dynamics of various education-related systems, such as organizational linkages among teachers and administrators, interrelationships among curricular modules, and psychological/cognitive processes of learning, to name a few. The three NetSciEd activities described above all aimed at producing all of those benefits, each to different degrees.

It goes without saying that the NetSciEd initiative is still far from completion, and there are many more goals that need to be accomplished in order to make major societal impacts. Urgently needed for teachers and students to adopt network science in their daily teaching and learning are resources such as: (1) detailed, modular, ready-to-use lesson plans that teachers can easily incorporate into existing curricula; (2) accessible, interactive online self-study guides for students; and (3) intuitive, interactive computer software tools for network drawing, visualization, analysis, and simulation. We are organizing a workshop for K-12 teachers to facilitate the development of lesson plans to produce (1) above (http://bit.ly/2s9GlRH), and we are in the process of exploring options for (2) and (3). Another major challenge in NetSciEd is how to better reflect the state-of-the-art of network science in K-12 educational materials, such as dynamical, temporal, adaptive nature of networks (Gross & Sayama 2009; Holme & Saramäki 2012; Porter & Gleeson 2016) and multiplex/multilayer/multiscale topologies of networks (Mucha et al. 2010; Kivelä et al. 2014; Boccaletti et al. 2014). While these concepts are highly applicable to many real-world networks, students and teachers typically find it quite difficult to understand dynamical behaviors or non-trivial topologies of networks. How to make these cutting-edge network science topics accessible and relevant to K-12 education remains an open challenge.

**Acknowledgments**

We thank Mason Porter and Ralucca Gera for their substantial contributions to the NetSciEd initiative. We also thank the following individuals and organizations for their help, support, and advice for NetSciEd:
Alvar Agusti, Suzanne Aleva, Aya Al-Zarka, Chris Arney, Albert-László Barabási, Rosa Benito, Robert F. Chen, Chantal Bonner Cherifi, Tara Chudoba, Kate Coronges, Albert Díaz, Raissa D'Souza, Brooke Foucault-Welles, Yoshi Fujiwara, Arthur Hjorth, Andreas Joseph, Przemysław Kazienko, Khaldoun Khashanah, Yasamin Khorramzadeh, Florian Klimm, Erik Laby, Mi Jin Lee, Sang Hoon Lee, Cheng-Te Li, Flora (Xianglin) Meng, Mor Nitzan, Peter Pollner, Elene Pugacheva, Jon Roginski, Gemma Rosell, Sarah Schoedinger, Rawan Shabbar, Saray Shai, Tarcizio Silva, H. Eugene Stanley, Craig Strang, Toshi Tanizawa, Paolo Tieri, Paul Trunfio, Lyubov Tupikina, Paul van der Cingel, Adam Wierzbicki, Robin Wilkins, Eri Yamamoto, Taha Yasseri, Tzu-Chi Yen, NetSci High students and teachers, the Network Science Society, NetSci/CompleNet conference organizers and symposium participants, and all of the members of the network science community who have contributed to and supported NetSciEd.

This material is based upon work supported in part by the U.S. National Science Foundation under Grant No. 1027752, 1139478, and 1509079; the Army Research Office; and BAE Systems.